\documentclass[aps,prx,reprint,superscriptaddress,amsmath,amssymb]{revtex4-2}

\usepackage{graphicx, xcolor}

\begin{document}

\title{Collective Noise Filtering in Complex Networks}

\author{Tingyu Zhao}
\affiliation{Department of Industrial Engineering and Management Sciences, Northwestern University, Evanston, Illinois 60208, USA}
\affiliation{NSF-Simons National Institute for Theory and Mathematics in Biology, Chicago, Illinois 60611, USA}

\author{István~A.~Kovács}
\email{istvan.kovacs@northwestern.edu}
\affiliation{Department of Physics and Astronomy, Northwestern University, Evanston, Illinois 60208, USA}
\affiliation{Department of Engineering Sciences and Applied Mathematics, Northwestern University, Evanston, Illinois 60208, USA}
\affiliation{Northwestern Institute on Complex Systems, Northwestern University, Evanston, Illinois 60208, USA}
\affiliation{NSF-Simons National Institute for Theory and Mathematics in Biology, Chicago, Illinois 60611, USA}

\date{\today}

\begin{abstract}
Complex networks are powerful representations of complex systems across scales and domains, and the field is experiencing unprecedented growth in data availability. However, real-world network data often suffer from noise, biases, and missing data in edge weights, which undermine the reliability of downstream network analyses.
Standard noise filtering approaches, whether treating individual edges one-by-one or assuming a uniform global noise level, are suboptimal, because in reality both signal and noise can be heterogeneous and correlated across multiple edges.
As a solution, we introduce the Network Wiener Filter, a principled method for collective edge-level noise filtering that leverages both network structure and noise characteristics, to reduce error in the observed edge weights and to infer missing edge weights. We demonstrate the broad practical efficacy of the Network Wiener Filter in two distinct settings, the genetic interaction network of the budding yeast \textit{S.~cerevisiae} and the Enron Corpus email network, noting compelling evidence of successful noise suppression in both applications. With the Network Wiener Filter, we advocate for a shift toward error-aware network science, one that embraces data imperfection as an inherent feature and learns to navigate it effectively.
\end{abstract}

\maketitle

Ever since the early development of graph theory, mathematicians have expanded its toolkit far beyond elegant puzzles, building increasingly sophisticated theories for reasoning about connectivity and structure~\cite{bollobas1998modern}. In recent decades, network science has carried these abstract ideas into the empirical world, using networks to represent complex systems across diverse disciplines, including, among others, social science~\cite{lazer2009computational}, biology~\cite{barabasi2004network}, transportation~\cite{barthelemy2011spatial}, climate science~\cite{donges2009complex}, finance and economics~\cite{acemoglu2015systemic}. Analyzing these complex systems from a network perspective has enabled researchers to uncover hidden patterns, capture emergent properties, and generate unique insights~\cite{barabasi2012network}.
While the network science community has benefited from a rapid expansion in the availability of network data, it is frequently overlooked that such data can be imperfect. Real-world networks are often subjected to various kinds of errors such as noise, biases, and incomplete observation~\cite{marsden1990network, krogan2006global, von2005string, costanzo2016global}. Yet it remains common in the literature to omit any quantitative assessment or reporting of these errors~\cite{newman2018networks}. Moreover, most network analysis pipelines do not consider potential data errors and treat empirical networks as if they were exact graphs, in the spirit of graph theory. Ignoring the issue of data reliability in this manner can compromise the robustness of downstream network-based inference, leading to flawed conclusions and decisions.

In response, a growing body of work has been seeking to mitigate edge noise in network data by exploiting network topological information, including link prediction methods~\cite{liben2003link, clauset2008hierarchical, zhou2009predicting, liu2010link, barzel2013network, lu2015toward, zhang2018link, kovacs2019network, wang2023assessment}, diffusion-based methods~\cite{feizi2013network, wang2018network, yu2023network}, generative model-based methods~\cite{butts2003network, guimera2009missing, newman2018network, newman2018estimating, peixoto2018reconstructing, peel2022statistical}, and singular value shrinkage methods~\cite{cai2010singular, benaych2012singular, shabalin2013reconstruction, nadakuditi2014optshrink, gavish2014optimal, josse2016adaptive, gavish2017optimal, thibeault2024low}. Each method category relies on distinct, often strong, network structural assumptions and is tailored to different use cases (SI Table I).
However, none of these approaches is adequate for the realistic scenario of having heterogeneous and correlated noise.
Either noise level is not explicitly incorporated, as in link prediction and diffusion-based methods, or noise is assumed to be largely homogeneous and independent across edges, as in generative model-based and singular value shrinkage methods.
In the case of heterogeneous, or even correlated, noise, practitioners often resort to local thresholding strategies at the level of individual edges, without harnessing the rich information in the network architecture~\cite{costanzo2016global}. This, in turn, can limit how effectively true signal can be extracted from the noisy data.

In this work, we propose the Network Wiener Filter (NetWF) as a principled method for collective noise filtering in complex networks.
Grounded in the classical theory of Wiener filtering~\cite{wiener1949extrapolation}, the NetWF takes into account the network structure through an intuitive edge-similarity measure, which is then combined with explicit edge-level noise statistics in a synergistic way.
Importantly, this formulation naturally accommodates heterogeneous noise or noise correlated across edges, as well as the special case of homogeneous and independent noise, rendering it fully general in terms of noise structure. At the same time, the NetWF applies seamlessly to directed or undirected, binary, weighted or even signed networks.
As demonstrated with our real-world applications, our efficient implementation using the conjugate gradient algorithm enables the NetWF to scale to large empirical networks.


\section*{From generalized Wiener filter to the NetWF}

In classical signal processing, the Wiener filter is a foundational framework designed to recover the underlying true signal from noisy observations by minimizing the expected mean squared error (MSE) between the estimate and the (unknown) true signal~\cite{wiener1949extrapolation, oppenheim2017signals}. Using the statistical properties of both the signal and the noise, it seeks an optimal balance between noise suppression and signal preservation. As a result, Wiener filtering has been widely adopted for denoising tasks such as in time series (Figure~\ref{fig:toy}A) and image processing (Figure~\ref{fig:toy}B).

\subsection*{Generalized Wiener filter}

Here, we start with a generalized Wiener filter formalism~\cite{pratt2006generalized} and extend it to network data.
Imagine that we observe a data vector $a = u + n \in \mathbb{R}^k$, where $u$ denotes the true signal and $n$ the additive noise; both of which are zero-mean random vectors drawn from unknown data generating processes. For data types that are not inherently one-dimensional, $a$ can be understood as a vectorized representation, such as a vectorized image or a vectorized network adjacency matrix, $a=\mathrm{vec}(A)$, where $A_{ij}$ stands for the interaction weight between nodes $i$ and $j$ in the network.
The goal is then to identify a ``Wiener operator" $G^{\mathrm{W}}$ that, when acting on $a$, gives an estimate of the signal $\hat{u} = G^\mathrm{W} a$ that minimizes the MSE with respect to the $u$:
\begin{align} \label{eqn:optimization}
G^{\mathrm{W}} = \, \arg\min_{G} \, \mathbb{E} \left \lVert Ga - u \right \rVert^2_2,
\end{align}
where the expectation $\mathbb{E}$ is taken over the joint realizations of $u$ and $n$, and $\lVert \cdot \rVert_2$ is the $L^2$ norm.
Assuming known second-order statistics of the data generating processes governing $u$ and $n$, the optimization problem of Eq.~\ref{eqn:optimization} admits a closed-form analytical solution; see SI Text B for a full derivation. Specifically, in the case when $u$ and $n$ are drawn from independent distributions, the solution is given by
\begin{align} \label{eqn:wf}
G^{\mathrm{W}} = C_u (C_u + C_n)^{-1},
\end{align}
in terms of the covariance matrices $C_u = \mathbb{E}[uu^T]$ and $C_n = \mathbb{E}[nn^T]$, both with dimensionality $\mathbb{R}^{k\times k}$, where the expectations are taken over the independent realizations of $u$ and $n$.

\begin{figure}[t!]
\centering
\includegraphics[width=1\linewidth]{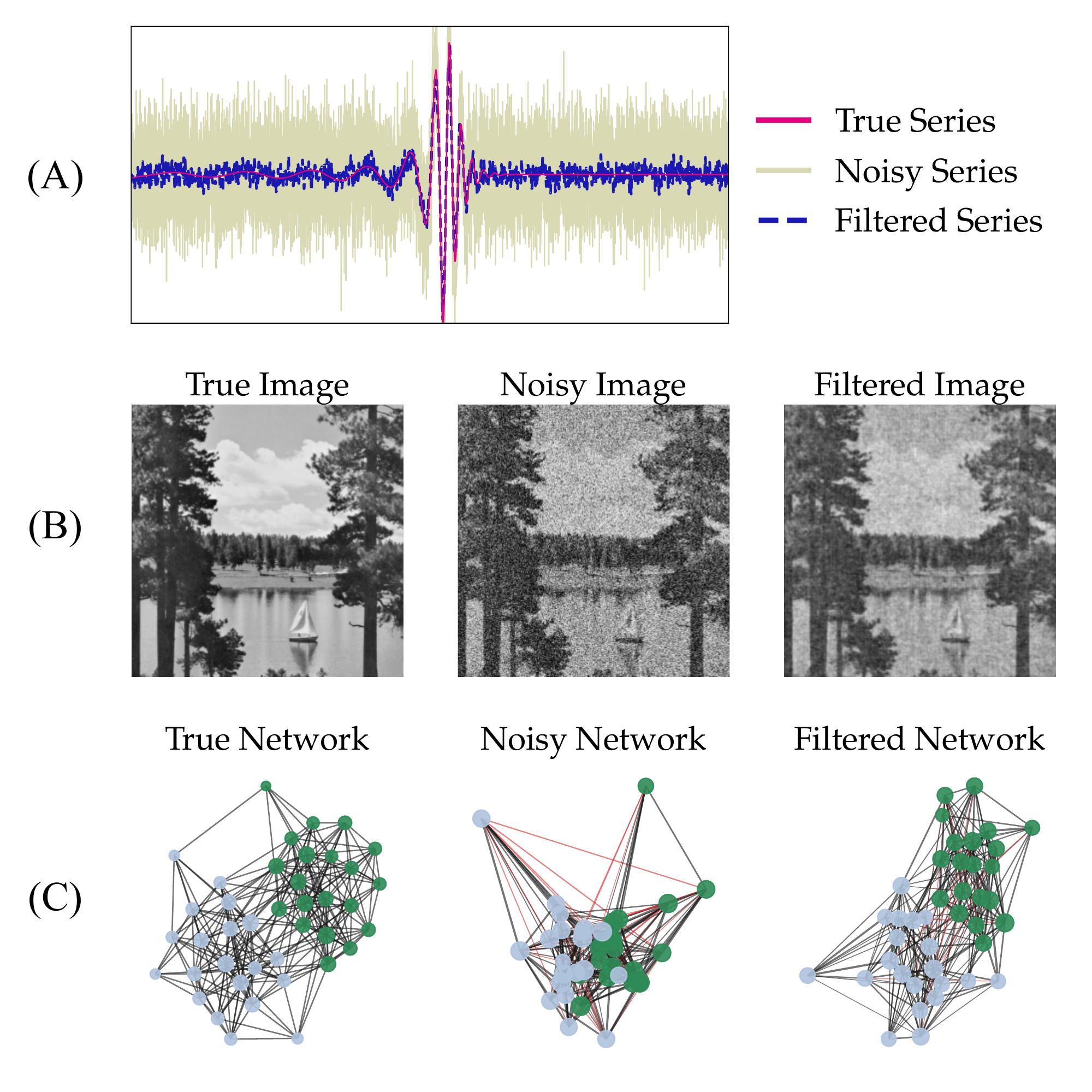}
    \caption{\textbf{Illustration of Wiener filtering for noise reduction across data types, and why networks are different.}
    Starting from (A) a gravitational wave time series, (B) an image, and (C) a toy network with two communities, we add independent and identically distributed Gaussian white noise to obtain noisy observations. We then apply Wiener filtering to estimate the underlying signal from noisy observations using the known noise statistics
    (more details in SI Text A).
    In time series and images, the classical Wiener filter exploits data-independent temporal or spatial proximity to capture signal correlations, with the intuition that nearby samples tend to be similar. In contrast, network data is inherently high-dimensional and lacks a built-in notion of proximity, requiring a new, data-adaptive treatment.
    The NetWF bridges this gap by using a network-informed edge-similarity measure that captures correlations globally across edges, as discussed in the main text. In the toy network, our NetWF reveals the original structure of two communities (more quantitative results in SI Figure 1).
    }
    \label{fig:toy}
\end{figure}

Applying the Wiener filter hinges on having access to the two covariance matrices, or, more realistically, reliable proxies for them. $C_n$ reflects noise characteristics and typically must be supplied through prior knowledge of how noise enters the data. In the simplest case, noise is independent and homogeneous, and $C_n$ reduces to a scaled identity matrix, as we imposed in our illustrative examples in Figure~\ref{fig:toy}. More generally, noise can be heterogeneous across entries, or even correlated, in which case $C_n$ is a generic, positive semidefinite covariance matrix.
On the other hand, $C_u$ boils down to signal variability and correlation, where the latter can be interpreted as the intrinsic similarity structure of the signal. For one-dimensional time series and two-dimensional images, this similarity structure is guided by proximity, as nearby samples are more similar than distant ones.
Note that all images share the same grid-like data architecture, independent of the actual pixel values.
Then it is straightforward to specify a similarity structure that respects this regular geometry, often \emph{a priori} by positing a stationary kernel.
It is also convenient to work in the frequency domain via the Fourier transform, where $C_u$ can be described compactly by a power spectral density through the Wiener–Khinchin theorem, achieving high computational efficiency for the Wiener filter~\cite{oppenheim2017signals}.

In contrast, networks (Figure~\ref{fig:toy}C) are fundamentally different: they are inherently high-dimensional objects whose intricate signal correlations must be inferred from the network data itself.
To provide the required context for Wiener filtering, we construct an edge-similarity measure that serves as a proxy for signal correlation in a qualitatively different, data-adaptive way, one that exploits the observed network patterns. 
In the proposed implementation of the NetWF, we do so using the intuition that edges between similar pairs of nodes are expected to be similar, as discussed next.
In the example of Figure~\ref{fig:toy}C, the NetWF successfully reveals the original structure of two network communities.

\subsection*{Network-informed edge-similarity measure}

\begin{figure}[t!]
\centering
\includegraphics[width=\linewidth]{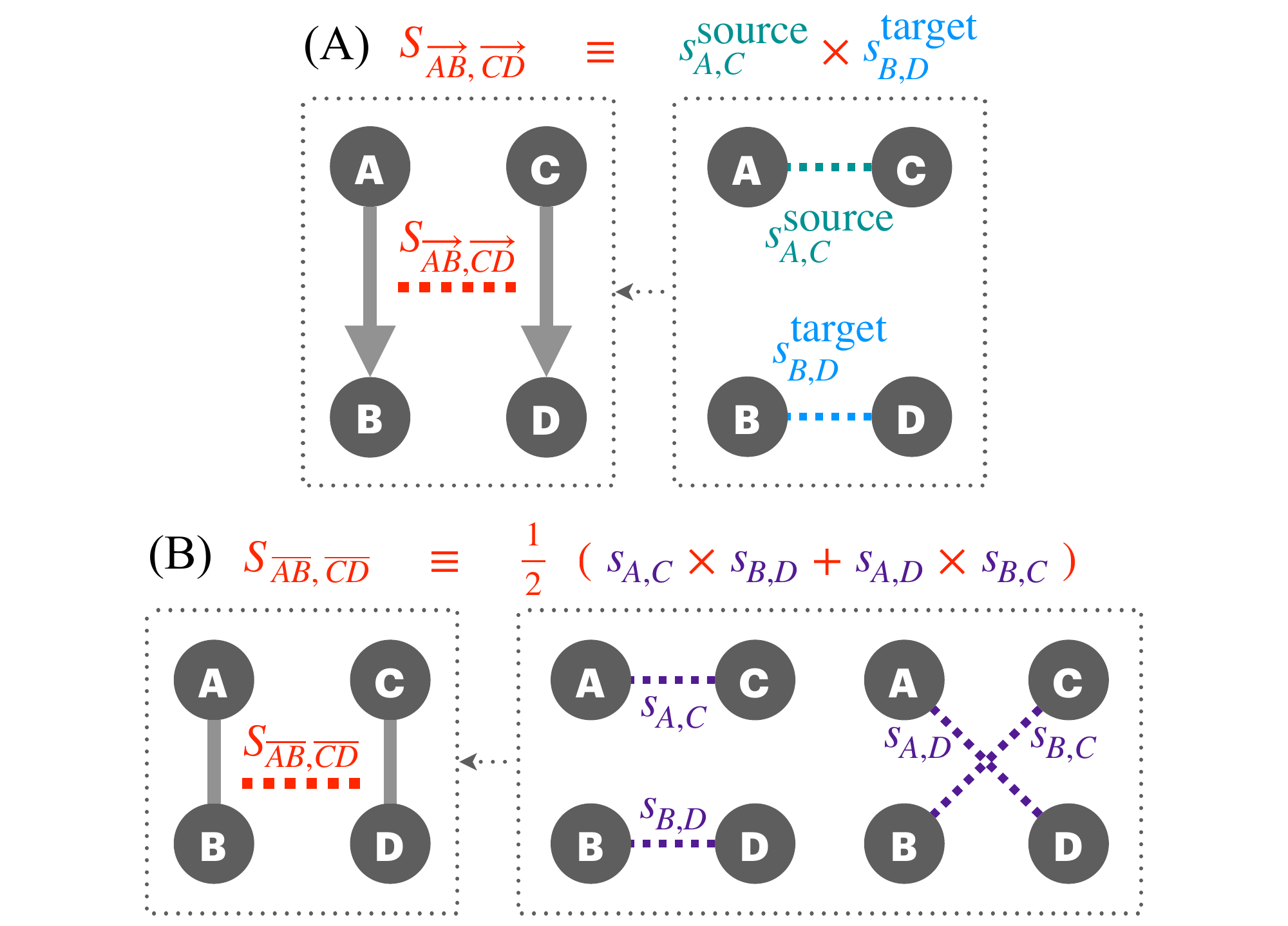}
    \caption{\textbf{Construction of our global edge-similarity measure, a key ingredient in the NetWF.} We propose that edges are similar if their endpoints are similar.
    The similarity between two nodes is in turn quantified by correlating their connection profiles with all other nodes in the network. (A) In a directed network, for edges $\overrightarrow{AB}$ and $\overrightarrow{CD}$, we separately consider the two sources ($A$ vs.\ $C$) using their outgoing connection patterns and the two targets ($B$ vs.\ $D$) using their incoming connection patterns (Eq.~\ref{eqn:directed_edge_sim}). (B) In an undirected network, the similarity between edges $\overline{AB}$ and $\overline{CD}$ is computed via considering the two configurations of endpoint matching, yielding a label-invariant similarity measure (Eq.~\ref{eqn:undirected_edge_sim}).
    }
    \label{fig:ansatz}
\end{figure}

For the purpose of community detection, a local notion of edge-similarity has already been proposed for edges that share an endpoint~\cite{ahn2010link}. The fundamental idea is that a suitable notion of node similarity can be used between the non-shared endpoints of the two edges to quantify the similarity of the edge pair. 
Here, we go one step beyond this local notion of edge-similarity and introduce a global edge-similarity between any two edges in the network, starting with the case of directed networks (Figure~\ref{fig:ansatz}A). For two directed edges $\overrightarrow{AB}$ and $\overrightarrow{CD}$, we separately model the similarity between the two source nodes, $A$ and $C$, and between the two target nodes, $B$ and $D$. As a convenient choice that works for any (binary, weighted or signed) networks, we quantify these node similarities using the Pearson correlation between their connection profiles. Specifically, the source profile similarity compares outgoing connection patterns,
\begin{align} \label{eqn:source_node_sim}
s^{\text{source}}_{i,j} 
\equiv \mathrm{corr}\!\left( w_{\overrightarrow{i\cdot}},\, w_{\overrightarrow{j\cdot}} \right),
\end{align}
where $w_{\overrightarrow{i\cdot}}$ collects the weights of edges pointing from node $i$ to all other nodes. Likewise, the target profile similarity compares incoming connection patterns,
\begin{align} \label{eqn:target_node_sim}
s^{\text{target}}_{i,j} 
\equiv \mathrm{corr}\!\left( w_{\overrightarrow{\cdot i}},\, w_{\overrightarrow{\cdot j}} \right),
\end{align}
where $w_{\overrightarrow{\cdot i}}$ collects the weights of edges pointing from all other nodes to node $i$.
Note that our NetWF framework is compatible with alternative notions of node similarity if desired, as also discussed later. 
As a last step, we then combine endpoint similarities to define the similarity between the directed edges $\overrightarrow{AB}$ and $\overrightarrow{CD}$ as
\begin{align} \label{eqn:directed_edge_sim}
S_{\,\overrightarrow{AB}, \,\overrightarrow{CD}}
\equiv s^{\text{source}}_{A,C}\, s^{\text{target}}_{B,D}.
\end{align}
For undirected networks (Figure~\ref{fig:ansatz}B), nodes no longer play distinct source or target roles, and the profile similarity network (PSN) between two nodes is defined simply as 
\begin{align} \label{eqn:undirected_node_sim}
s_{i,j} 
\equiv \mathrm{corr}\!\left( w_{\overline{i\cdot}},\, w_{\overline{j\cdot}} \right),
\end{align}
where $w_{\overline{i\cdot}}$ collects the weights of edges attached to node $i$. We obtain the similarity between the edge pair $\overline{AB}$ and $\overline{CD}$ by averaging over both configurations of endpoint matchings,
\begin{align} \label{eqn:undirected_edge_sim}
S_{\,\overline{AB}, \,\overline{CD}}
\equiv \frac{1}{2}\!\left(
s_{A,C}\, s_{B,D}
+ s_{A,D}\, s_{B,C}
\right).
\end{align}
Although the Wiener filter appears to be a linear framework with respect to input data, the filtering process is actually nonlinear if the Wiener operator $G^{\mathrm{W}}$ of Eq.~\ref{eqn:wf} itself is data-dependent, as is the case for the proposed NetWF. 
For detailed computational implementation of the NetWF, including a conjugate gradient-based iterative implementation that makes handling large empirical networks feasible, see Materials and Methods.

\section*{Applications in real-world complex networks}

Next, we apply the NetWF to two real-world complex networks to assess its practical performance in two distinct settings: biological vs.~social system, experimental vs.~sampling noise, static vs.~dynamic network, and undirected vs.~directed edges.
For comparison, we select the Optimal Shrinker (OS) of singular values as a state-of-the-art baseline noise filtering method~\cite{josse2016adaptive}. The OS performs spectral compression by assuming that the adjacency matrix has low effective rank~\cite{thibeault2024low}. Although it applies broadly across network types, it is formulated for homogeneous noise only, so we need to homogenize heterogeneous noise in the input before applying the OS. See Materials and Methods for more details.

\subsection*{Biological network application: yeast genetic interaction network}

Large-scale biological data are inherently noisy and often incomplete, 
yet we rarely have detailed, quantitative insights into how much error there is in the measurements. An exception is the genome-wide 
pairwise genetic interaction (GI) network of the budding yeast \textit{S.~cerevisiae}~\cite{costanzo2016global}.
Pairwise GIs arise when the fitness phenotype of a double mutant, $f_{i,j}$, deviates from the expectation under independent single-mutant effects, $f_i$ and $f_j$:
\begin{align} \label{eqn:gi}
    \epsilon_{i,j} = f_{i,j}  - f_i f_j\;.
\end{align}
Here, the fitness is measured as colony growth relative to a wild-type (non-mutant) reference under standardized conditions, via high-throughput synthetic genetic array (SGA) analysis~\cite{baryshnikova2010quantitative}.
Notably, the experiments not only provide the $\epsilon_{i,j}$ edge weights for each gene pair, but also the experimental error bar for each weight~\cite{costanzo2016global}. Most interaction weights were found to be comparable to the noise level, suggesting that the independent fitness expectation is valid for most gene pairs.
A negative GI ($\epsilon_{i,j}<0$) indicates that the double mutant grows worse than expected, in extreme cases known as synthetic lethality. Conversely, a positive GI ($\epsilon_{i,j}>0$) indicates that the double mutant grows better than expected, i.e.~genetic buffering or suppression.
To identify true GIs, the authors proposed three thresholding schemes (lenient, intermediate, stringent) for edge-level data based on a combination of effect size ($\epsilon_{i,j}$) and noise reporting ($p$-value). 

These systematic fitness screens uncovered a rich hierarchical functional organization in an eukaryotic cell, covering most of the pairwise GI space between $\sim 6{,}000$ genes in \textit{S.~cerevisiae}. The organization of GIs was found to be qualitatively different for essential genes (required for growth) vs.~non-essential genes, as well as for positive vs.~negative GIs. 
Here, we focus our analysis on the dense part of the network between essential genes (ExE), comprising $855$ nodes; see Materials and Methods for details. 
Besides providing rich functional information, it was also found that GIs encode additional biological information~\cite{costanzo2016global}. In the ExE space, this is manifested in the significant overlap between negative GIs and other data sources, including protein--protein interactions (PPI)~\cite{oughtred2021biogrid}, protein co-complex membership~\cite{baryshnikova2010quantitative,benschop2010consensus}, and Gene Ontology (GO) bioprocess term co-annotation~\cite{cherry2012saccharomyces}; see Materials and Methods for details. The relatively few cases of extreme positive GIs also provide information on genetic function; however, positive GIs were deemed to be somewhat problematic, potentially being diluted with false positives, prompting additional focused experimental efforts on genetic suppression~\cite{van2016exploring, van2020systematic}.

Using a stringent weight threshold~\cite{costanzo2016global}, $\epsilon<-0.12$ (negative) and $\epsilon>0.16$ (positive), the ExE map contains $3{,}516$ positive GIs and $26{,}268$ negative GIs. The noise variance information entails uncorrelated but heterogeneous  edge-level noise, corresponding to a diagonal noise covariance matrix $C_n$. In contrast, applying the NetWF denoising (SI Figure 2) and then applying the same stringent weight threshold yields only $620$ positive GIs ($82\%$ reduction in count) and $17{,}558$ negative GIs ($33\%$ reduction in count), including $2034$ new negative GIs.
In other words, the NetWF disproportionately reduces the number of (potentially problematic) positive GIs, while even unveiling previously neglected negative GIs.

\begin{figure}[t!]
\centering
\includegraphics[width=\linewidth]{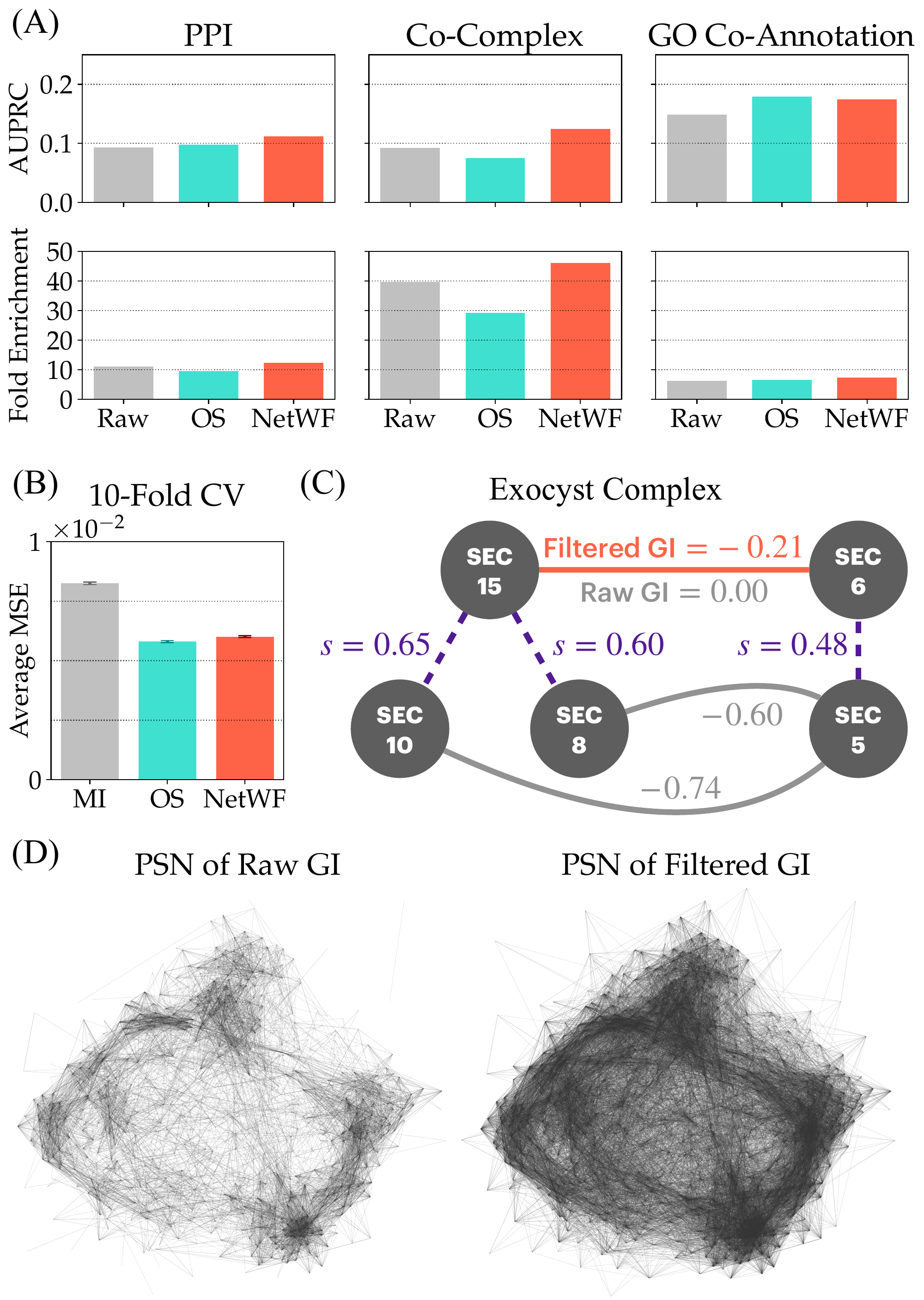}
    \caption{\textbf{Evidence that the NetWF reduces experimental noise in the budding yeast \textit{S.~cerevisiae} genetic interaction (GI) network}.
    (A) Validation of negative GI values against three biological benchmarks: protein--protein interaction (PPI), protein co-complex membership, and Gene Ontology (GO) bioprocess term co-annotation. Relative to the raw data, NetWF achieves higher area under the precision--recall curve (AUPRC) and larger fold enrichment across all benchmarks, while the optimal shrinker (OS) baseline delivers inconsistent performance.
    (B) Ten-fold cross validation (CV) test: in each fold, a random subset of GI entries is masked and predicted from the remaining data. The NetWF and the OS achieve significantly lower average mean squared error (MSE) than na\"ive mean imputation (MI), where the missing entries are imputed using the mean of available GI values in each gene. Error bars indicate standard errors across folds.
    (C) Example of inferring a novel GI within the exocyst complex (SEC15--SEC6) by the NetWF. 
    Dashed edges highlight a few example profile similarities within the complex that facilitate information propagation, collectively shifting the GI value from zero to a strongly negative estimate.
    (D) Profile similarity networks (PSNs) showcasing similarities between gene pairs (Eq.~\ref{eqn:undirected_node_sim}), before and after the NetWF is applied to the raw GI network. Edge width scales with similarity, and similarities below $0.2$ are omitted for clarity. 
    }
    \label{fig:Yeast}
\end{figure}

More importantly, the filtered ExE network demonstrates stronger adherence to all three benchmarks (PPI, co-complex, GO bioprocess term co-annotation), improving both area under the precision--recall curve (AUPRC) and fold enrichment relative to the raw data (Figure~\ref{fig:Yeast}A). Here, we compute AUPRC by ranking gene pairs from the most negative GI upward and synthesizing how effectively benchmark pairs are recovered as the selection threshold is progressively relaxed. Fold enrichment measures how overrepresented benchmark pairs are among the $1{,}000$ most negative GIs compared with their background prevalence in the overall ExE space.

Next, to examine link prediction performance, we carry out a ten-fold cross validation test (Figure~\ref{fig:Yeast}B). We divide the observed symmetric GI entries into 10 partitions, and in each fold, we mask the partitioned entries to missing values and infer them from the remaining observations; see Materials and Methods for details.
The NetWF delivers substantially lower MSE than a na\"ive mean imputation that fills missing entries using the mean of available GI values for each gene.

As a case study, we successfully recover a missing interaction from the input data within the exocyst complex~\cite{mei2018cryo}. The NetWF infers a strong negative GI between genes SEC15--SEC6 ($\epsilon = -0.21$), in agreement with the known biophysical interaction~\cite{gavin2006proteome,songer2009sec6p}. Although the prediction originates from the NetWF processing collective information from the entire network, we can provide an interpretation within the exocyst subgraph (Figure~\ref{fig:Yeast}C). Notably, SEC15 is highly similar to SEC10 and SEC8 in GI profile, while SEC6 is similar to SEC5; therefore, the strong negative GI of SEC10--SEC5 and SEC8--SEC5 contribute to the negative SEC15--SEC6 estimate. See SI Figure 3 for the physical structure of the exocyst complex, noting the SEC15--SEC6 physical contact.

Finally, to quantify gene-level similarity, we construct PSNs in which edge weights are Pearson correlations between GI profiles (Eq.~\ref{eqn:undirected_node_sim}). Using the originally proposed cutoff $>0.2$~\cite{costanzo2016global}, the PSN has $7{,}563$ edges before filtering and $22{,}685$ edges after filtering (Figure~\ref{fig:Yeast}D). This three-fold increase indicates fewer idiosyncratic differences in gene connection profiles, i.e.~a smaller effective degrees of freedom. Together with the evidence that the biological signal is preserved or even enhanced (Figure~\ref{fig:Yeast}A), this suggests a higher signal-to-noise ratio after filtering.

In comparison, the OS baseline behaves inconsistently across evaluations. It adheres worse to the protein co-complex membership benchmarks than the raw data (Figure~\ref{fig:Yeast}A), yet performs well in cross validation (Figure~\ref{fig:Yeast}B), yields a dense PSN ($58{,}630$ edges), and also recovers the SEC15--SEC6 interaction ($\epsilon = -0.15$).

To test how much the performance of the NetWF depends on having access to the heterogeneous noise information, we run a restricted, homogenized-noise version in which we replace the heterogeneous noise variances by their mean, just like for the OS (SI Figure 4).
The restricted results across the board change only marginally compared to those with the full NetWF, with over a $98\%$ overlap among the negative GIs. This result indicates that the NetWF can be a useful tool across various applications, even without detailed noise characterization, a point we further investigate in our next application.

\subsection*{Social network application: the Enron Corpus email network}

The Enron Corpus is a publicly available database containing internal email exchanges among key Enron employees in the years leading up to the company's famous bankruptcy in December 2001, and it has become a widely used resource for studying organizational communication patterns~\cite{klimt2004enron}.
We use this dataset for two complementary goals.
First, we demonstrate the performance of the NetWF in the case of heterogeneous and correlated noise covariance $C_n$, the highest level of noise characteristics information.
Second, we take the opportunity to embrace the more common empirical setting where no noise statistics is provided, and stress test the extent to which the NetWF can perform denoising under this far-from-optimal condition.

\begin{figure*}[t!]
\centering
\includegraphics[width=\linewidth]{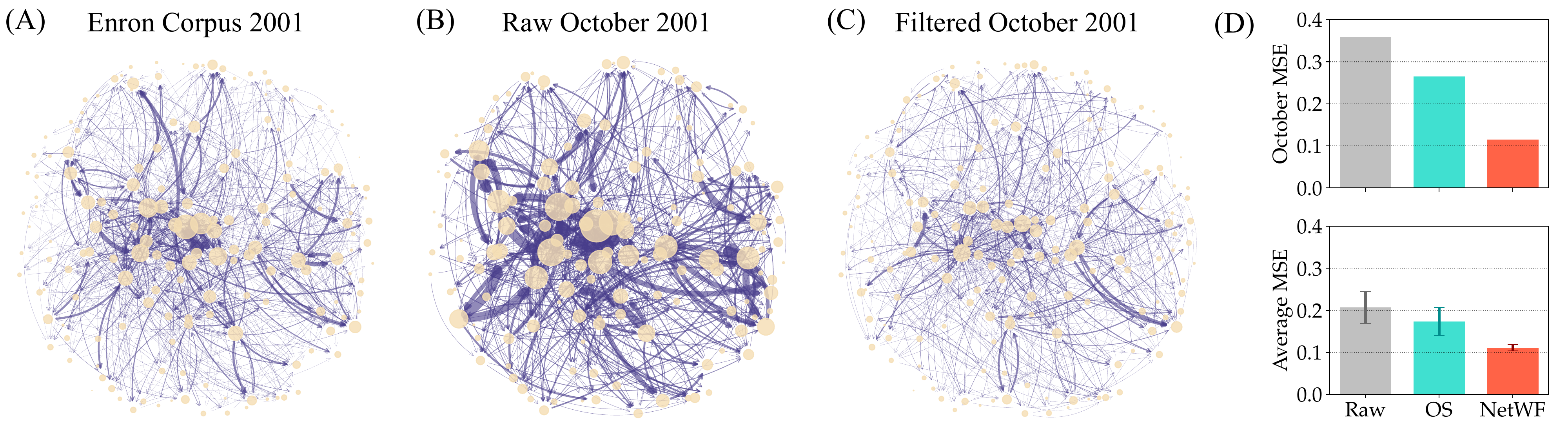}
    \caption{\textbf{Evidence that the NetWF reduces sampling noise in the Enron Corpus email network.} 
    (A) Full-year email exchange frequency network of 2001 between employees as nodes. Directed edges point from sender to recipient, with edge weight equal to email frequency (defined as the number of emails per month). Node size reflects the sum of incoming and outgoing weights. For clarity, edges with weight $<0.2$ are not shown.
    (B) Raw October 2001 snapshot, treated as a noisy calendar-month sample of the full-year network, exhibiting pronounced fluctuations in edge weights.
    (C) October 2001 after applying the NetWF, with the extreme weights reduced without compromising the overall structure.
    (D) Relative to the raw October network and the OS baseline, the NetWF largely reduces the October MSE with respect to the full-year network. Repeating for all twelve calendar-month snapshots, the NetWF also attains the lowest and most stable average MSE. Error bars indicate standard errors across calendar months.
    }
    \label{fig:Enron}
\end{figure*}

We focus on the year 2001, which spans the peak of the crisis. After removing self-emails, the dataset contains 12{,}684 time-stamped emails among 163 employees. We construct the full-year network as a weighted, directed email exchange frequency network, where each directed edge weight is the email frequency from sender to recipient (Figure~\ref{fig:Enron}A). The resulting network exhibits a clear core--periphery organization, suggesting complex social structure~\cite{csermely2013structure}.
The full-year network is a temporal aggregate of a dynamic communication process. To examine temporal variability, we also construct the twelve calendar-month snapshots of 2001 (SI Figure 5), each of which can be viewed as a noisy sample of the underlying full-year network, where the finite observation window causes sampling fluctuations in edge weights. An interesting denoising task is therefore to filter each calendar-month snapshot to see if that brings it closer to the full-year network, which we use as a reference signal.

To address the first goal, we use month-to-month fluctuations to estimate edge-level noise covariance and incorporate this information into the NetWF, producing twelve denoised calendar-month snapshots (SI Figure 6). For the OS baseline, only the mean variance over nodes calculated from the full noise covariance is used (SI Figure 7).
October 2001 provides an illustrative example because of its historical significance: Enron’s accounting fraud came to light, sharply accelerating the company’s downfall. Relative to the full-year network, the raw October snapshot shows turmoil with pronounced sampling fluctuations, including the presence of unusually strong edges (Figure~\ref{fig:Enron}B). After applying the NetWF, these extremes are smoothed while the overall structure remains preserved (Figure~\ref{fig:Enron}C).
Globally, in terms of the MSE with respect to the full-year network, the NetWF again outperforms both the raw snapshot and the OS. Over all twelve months, the NetWF achieves an average MSE of $0.11 \pm 0.01$, lower and more robust compared to the raw snapshots ($0.21 \pm 0.04$) and the OS outcomes ($0.17 \pm 0.03)$, as seen in Figure~\ref{fig:Enron}D.
Similar evaluations carried out for each individual calendar-month snapshot, including both MSE and $R^2$ perspectives, are shown in SI Figure 8. There, we also test the restricted version of the NetWF that uses homogenized noise information, in line with the OS.
Paired $t$-tests show that the full NetWF reduces MSE significantly relative to both the raw snapshots ($p = 0.01$) and the OS ($p = 0.03$). The restricted NetWF is still significantly better than the raw snapshots ($p = 0.02$), although its improvement over the OS is not significant ($p = 0.10$).

Upon further testing the NetWF, we also found an interesting directional asymmetry in this network. If we modify Eq.~\ref{eqn:directed_edge_sim} by using source-node similarity in place of target-node similarity, i.e.~defining $S_{\,\overrightarrow{AB}, \,\overrightarrow{CD}}
\equiv s^{\text{source}}_{A,C}\, s^{\text{source}}_{B,D}$, the NetWF delivers even stronger denoising results with an average MSE of $0.07 \pm 0.01$, as shown in SI Figure 9. This suggests that similarity in sending behaviors might carry more information about the edge weight correlations in this network.
More broadly, this finding highlights that although our proposed edge-similarity measure is generically applicable, even better performance can be achieved if the similarity measure is tailored to the structure of the specific network dataset. 

Last, we turn to our second goal with this dataset, that is, denoising when $C_n$ is unknown. In this setting, we use a na\"ive guess of noise level for both the NetWF and the OS, using the variance across edge weights in the observed single calendar-month snapshot as the homogeneous noise variance.
Even under this under-informed condition, the NetWF achieves significantly lower average MSE ($0.14 \pm 0.01$) compared to both the raw snapshots and the OS baseline ($0.18 \pm 0.03$) according to paired $t$-tests, with $p$-values $0.02$ and $0.04$ respectively. Therefore, successful denoising is possible with the NetWF even with minimal inputs, as long as a reasonable noise level is assumed; see SI Figure 10.


\section*{Discussion}

Wiener filtering was the first example of a statistically solid, collective noise reduction technique~\cite{wiener1949extrapolation}. The generalized Wiener filter framework is mathematically grounded, as the Wiener operator (Eq.~\ref{eqn:wf}) optimally solves the MSE minimization problem (Eq.~\ref{eqn:optimization}). 
Yet, an extension to even static network data remained an unsolved problem until now. 
Our proposed NetWF offers a qualitative advancement in addressing the often-overlooked issue of noise in static networks. It is a versatile denoising method applicable to both directed and undirected networks, with arbitrary weights, including signed networks. 
A key to this advancement is our proposed concept of edge similarity, a generic solution that is intuitive and interpretable, building upon any preferred notion of node similarity. 

An exciting future direction is to utilize the NetWF to identify the most suitable measure of node (or edge) similarity adaptively in each dataset that optimizes the output of the NetWF. As investigated in our Enron study, such notions of similarity that are tailored to each application domain are expected to lead to further improved results. Another promising avenue is to use dimension reduction and machine learning techniques to learn the optimal similarity structure, such as using node2vec to generate low-dimensional node embeddings, which can then be used to model node (and edge) similarities~\cite{grover2016node2vec}.
The NetWF naturally adapts to different levels of uncertainty in network data; see SI Figure 11. Although the NetWF is expected to perform best when a full noise characterization is available, we have seen that it can deliver similarly good results even with much less information, such as a homogeneous noise estimate.

As large-scale biological data are inherently imperfect, this is a key area of application for the NetWF. As an illustration, we used a biologically coherent ExE portion of the entire SGA GI dataset for baker's yeast, where the strong structure of the network offered multiple reliable evaluations against benchmark datasets. A natural next step is to apply the NetWF to the entire yeast dataset as well as to similar datasets in other organisms, such as the recent genome-scale screen in human HAP1 cells~\cite{billmann2025global}.
By reducing data noise, these applications could further clarify the amount of agreement in the functional architecture across various organisms.

As showcased with the Enron Corpus email network study, the static NetWF is also applicable to snapshots of temporal networks. An opportunity for future development is a temporal NetWF that considers the full time series of network data, combining temporal Wiener filtering ideas with the NetWF. Further extensions include multiplex and multilayer networks~\cite{boccaletti2014structure, kivela2014multilayer}, as well as hypergraphs and simplicial complexes~\cite{battiston2020networks, bianconi2021higher}. Such higher-order interactions play an important role in genetics, as evidenced by the prevalence of GIs between sets of three genes in yeast~\cite{kuzmin2018systematic, kuzmin2021tau}, and are also increasingly studied in computational social science~\cite{shi2015weaving, lungeanu2021team}.

Just like in any form of quantitative data analysis, understanding and reporting uncertainties in networks is crucial, as these uncertainties can be leveraged to improve the overall quality and reliability of network-based inference. We believe that the NetWF serves as an important milestone towards a paradigm shift in network science, where reporting and utilizing noise characteristics for network data is seen as a valuable and integral part of understanding the underlying complex networks. 

\section*{Materials and Methods} 

\subsection*{Implementation of the NetWF}

We consider an observed network on $v$ nodes, represented by a demeaned adjacency matrix
$A\in\mathbb{R}^{v\times v}$ and its vectorization $a=\mathrm{vec}(A)\in\mathbb{R}^{k}$ with $k=v^{2}$, together with a noise covariance matrix $C_n\in\mathbb{R}^{k\times k}$ characterizing noise in edge weights.
We first compute node-level profile similarity networks (PSNs), symmetric matrices whose elements are similarities between node pairs, as in Eqs.~\ref{eqn:source_node_sim}--\ref{eqn:target_node_sim} for directed networks or Eq.~\ref{eqn:undirected_node_sim} for undirected networks. 
We present two NetWF implementations: a direct NetWF and an iterative NetWF. 

In the direct implementation of the NetWF, we first construct a network-informed signal covariance $C_u\in\mathbb{R}^{k\times k}$ from the observed topology. $C_u$ is directly calculated from the PSNs using the matrix Kronecker product, following the exact protocols in Eqs.~\ref{eqn:directed_edge_sim}~and~\ref{eqn:undirected_edge_sim}.
We then set the overall signal variability in $C_u$ by giving it a prefactor equal to the data variance across entries in $A$, thereby completing the construction of $C_u$.
We break down the signal estimate computation $\hat{u}=G^{\mathrm{W}}a = C_u (C_u + C_n)^{-1} a \equiv C_u x$ into two steps, avoiding explicit matrix inverse in Eq.~\ref{eqn:wf} for for numerical stability.
(i) solve the linear subproblem $(C_u + C_n + \epsilon I)x = a$ for $x\in\mathbb{R}^k$, where $\epsilon$ is a small positive number to assure that the matrix is numerically invertible and well-conditioned, providing a stable alternative to a Moore-Penrose pseudoinverse; 
(ii) compute $\hat{u} = C_u x$.
Finally, we reshape $\hat{u}$ to arrive at the denoised adjacency matrix $\hat{U} = \mathrm{unvec}(\hat{u})$.
Note that this direct implementation quickly becomes prohibitive for large complex networks, as storing $C_u$ and $C_n$ requires $O(v^{4})$ memory.

Hence, we develop an alternative, iterative implementation of the NetWF, which reduces the memory burden to $O(v^{2})$, as long as $C_n$ is diagonal (i.e.~noise is treated as uncorrelated across edges). The key is to avoid explicitly forming $C_u$.
Since $C_u+C_n$ is symmetric and positive semidefinite, we deploy the conjugate gradient (CG) algorithm to efficiently solve the linear subproblem $(C_u + C_n + \epsilon I)x=a$ with guaranteed numerical convergence~\cite{golub2013matrix}.
Note that CG does not require explicit formation of $C_u+C_n$, but only its action on vectors. Consequently, in both the CG subproblem and the subsequent $C_u x$ computation, we exploit the Kronecker-product structure of $C_u$ via the identity $(M_1 \otimes M_2) x = \mathrm{vec}\left(M_2 \, \mathrm{unvec}(x) \, M_1^{\mathsf T} \right)$, reducing the dimensionality of the problem from $\mathbb{R}^{k \times k}$ to $\mathbb{R}^{v \times v}$. This step resolves the memory bottleneck and allows the NetWF to scale to large empirical systems.

\subsection*{Singular value shrinkage}

As a baseline network denoising technique, we implement the optimal shrinker (OS) of singular values developed by Gavish and Donoho~\cite{gavish2017optimal}. The OS adopts the spectral theory of random matrices~\cite{benaych2012singular} and works under the assumption that the rank of the signal adjacency matrix is much smaller than its dimension. Conceptually, the OS procedure yields a denoised network by preserving the dominant global network patterns and suppressing fluctuations that are more likely to be noise.
For a $v$-by-$v$ matrix observed in homogeneous white noise of variance $\sigma^2$, the OS performs singular value decomposition and shrinks each singular value $y_i$ to 
\begin{align} \label{eqn:shrinkage}
y^*_i = 
\begin{cases}
\sqrt{y_i^2-4v\sigma^2}, & y_i \geq 2\sqrt{v}\sigma,\\
0, & y_i < 2\sqrt{v}\sigma,
\end{cases}
\end{align}
and the denoised matrix is obtained.
If we are given heterogeneous edge-level noise as an input, we calculate the mean of the noise variances and use it as the $\sigma^2$ input.

\subsection*{Data processing}

The \textit{S.~cerevisiae} GI network and its associated noise variance matrix were constructed from the SGA dataset reported in Ref.~\cite{costanzo2016global} and made available at \url{https://thecellmap.org/yeast/costanzo2016/}.
We study the densely connected essential GI network (ExE), between essential genes.
The dataset provides GI scores, double-mutant fitness standard deviations, and $p$-values for assays between $1{,}108$ query strains (covering $796$ query genes) and $792$ array strains (covering $561$ array genes). For query--array gene pairs with duplicated assays, we select the assay with the most significant $p$-value and drop the others.
We then take the union of unique query and array genes, yielding $855$ essential genes, and assemble a square GI matrix indexed by this gene set. 
If two experiments map to the same (undirected) gene pair, we average their $\epsilon$ values. This procedure produces a symmetric GI matrix representing an undirected, weighted network on the $855$ essential gene nodes.
The noise variance matrix is constructed analogously: for each assay we square the reported double-mutant fitness standard deviation $\sigma$ to obtain an edge-level variance $\sigma^2$, place this value into the same $855\times 855$ gene-indexed matrix, and, when multiple assays map to the same (undirected) gene pair, average their $\sigma^2$ values.

Note that the raw ExE matrix, as well as the corresponding variances, contains missing entries. 
During cross validation, we mask more entries to create additional missing values.
Before ingesting the data into the NetWF (including PSN computation) or the OS, we impute missing noise variances with the sum of the mean of observed variances over entries in the noise variance matrix and the variance of observed GI scores over entries in the ExE matrix, and then impute missing GI scores with the mean of observed GI scores over entries in the ExE matrix.

The PPI benchmark network was obtained from BioGRID~\cite{oughtred2021biogrid}, version 5.0.253 (compiled on Dec.~25, 2025), at \url{https://thebiogrid.org}, filtering interaction types to retain only physical association and direct interaction.
The protein co-complex membership benchmark network was derived from the protein complex standard provided by Ref.~\cite{costanzo2016global} based on Refs.~\cite{baryshnikova2010quantitative,benschop2010consensus}.
The Gene Ontology (GO) bioprocess term co-annotation benchmark network was constructed following the protocol in Ref.~\cite{costanzo2016global}, with the GO bioprocess term functional annotation data downloaded from the Saccharomyces Genome Database (SGD)~\cite{cherry2012saccharomyces} at \url{https://www.yeastgenome.org}.

The Enron Corpus email network was obtained from \url{https://www.cs.cmu.edu/~enron/}, using the latest May~7, 2015 release.
Both the NetWF and the OS baseline can introduce self-links and negative edge weights, even when the input network contains no self-links or negative edge weights. For visualization and evaluation in the Enron application, we therefore remove self-links and truncate negative weights to zero for all denoised outputs.

\subsection*{Data and code availability}
For full reproducibility of the NetWF results in this study, an archived release of relevant code and processed data is available on Zenodo: \href{https://doi.org/10.5281/zenodo.18472712}{10.5281/zenodo.18472712}. The development repository is maintained at \url{https://github.com/markzhao98/NetWF_paper}.


\section*{Acknowledgments} 

The authors acknowledge funding through the `CAREER: Network-based inference of complex biological interactions' PHY-2440223 Physics of Living Systems (POLS) NSF CAREER Award, sponsored by the NSF 22-586 Faculty Early Career Development Program. We thank Bingjie Hao, Ruiting Xie, Anastasiya Salova, Eli Daniel Ganz, Leone Luzzatto, Maryn Carlson, Noshir Contractor, Michael Costanzo, and Charles Boone for their helpful comments and discussion. 
T.Z. and I.A.K. acknowledge support from the NSF–Simons National Institute for Theory and Mathematics in Biology, jointly funded by the U.S. National Science Foundation (Award No. 2235451) and the Simons Foundation (Award No. MP-TMPS-00005320).

\bibliography{WF_ref}

\clearpage

\end{document}